\documentclass[epjST]{svjour}
\usepackage{graphics}
\usepackage{epsfig,graphicx,tabularx,caption,subcaption}
\usepackage{color}
\usepackage{subfig}
\begin{document}
\title{Non-Local Order Parameters as a Probe for Phase Transitions in the Extended-Fermi Hubbard Model}
\author{Luca Barbiero\inst{1,2} \and Serena Fazzini\inst{3} \and Arianna Montorsi\inst{3}}
\institute{CNR-IOM DEMOCRITOS Simulation Center and SISSA, Via Bonomea 265, I-34136 Trieste, Italy \and Dipartimento di Fisica e Astronomia "Galileo Galilei", Universit\`a di Padova, 35131 Padova, Italy \and Institute for condensed matter physics and complex systems, DISAT, Politecnico di Torino, I-10129, Italy}
\abstract{The Extended Fermi-Hubbard model is a rather studied Hamiltonian due to both its many applications and a rich phase diagram. Here we prove that all the phase transitions encoded in its one dimensional version are detectable via non-local operators related to charge and spin fluctuations. The main advantage in using them is that, in contrast to usual local operators, their asymptotic average value is finite only in the appropriate gapped phases. This makes them powerful and accurate probes to detect quantum phase transitions. Our results indeed confirm that they are able to properly capture both the nature and the location of the transitions. Relevantly, this happens also for conducting phases with a spin gap, thus providing an order parameter for the identification of superconducting and paired superfluid phases.}
%
\maketitle
\section{Introduction}
\label{intro}
The one-dimensional (1D) extended Fermi-Hubbard Model (EFM) is a lattice Hamiltonian describing a hopping fermionic mixture subject to both density-density nearest-neighbor (NN) and on-site interaction. It has a huge range of applications being able to properly describe copper-oxide materials related to the high-$T_c$ cuprate superconductors \cite{emery}, conducting polymers \cite{polimers}, and organic charge-transfer salts \cite{ishiguro}.\\
The full EFM phase diagram has been studied for specific fillings by means of different techniques \cite{Hir,LiHi,CaFr,PeMi,nakamura,lemini}. In particular, the repulsive interaction regime has been intensively investigated due to the presence of a phase with bond-order waves (BOW) \cite{nakamura}, not predicted within single loop bosonization \cite{Tsuchiizu1}. Recently it has been shown that non-local order parameters (NLOPs) are able to give an accurate description of the latter regime \cite{BMR} as well as in presence of density unbalance \cite{dhar}.\\
Relevantly, in the field of ultracold quantum gases \cite{Bloch2008}, the bosonic counterpart of the EFM with dipole-dipole interaction has been experimentally realized by means of $Er$ magnetic atoms \cite{baier}. Furthermore several kinds of particles with Fermi statistics and strong dipolar momentum \cite{lu,aikawa,park} are currently available making just matter of time the experimental realization of the EFM Hamiltonian. Motivated by this aspect, theorists have studied the EFM with long-range dipole interaction \cite{DBRD} showing that the phase diagram is basically captured also by considering only NN interaction. Furthermore in-situ imaging has allowed for the experimental detection of a specific type of NLOP \cite{endres} while the other ones are in principle measurable with the ongoing experimental techniques \cite{parsons}.\\
The aforementioned experimental achievements motivated us to investigate all the phase transitions occurring in the EFM by means of  the NLOPs since, as conjectured in \cite{BMR}, they should be able to properly capture all the possible gapped phases. Further crucial reasons to use them are: they capture the presence of non-local long range order, which does not break any continuous symmetry, thus being order parameters in 1D system without violating Mermin-Wagner theorem \cite{mermin}; they can signal the presence of non-trivial topological phases \cite{MDIR}; they can detect hidden magnetism not captured by two-points correlation functions \cite{dennijs}.\\
In this paper, after a review of the model and its 1D phases, we introduce the different NLOPs describing their implications on virtual excitations, magnetic and possible topological orders. In section 4 we present a density matrix renormalization group (DMRG) \cite{white} analysis of the behavior of the NLOPs at each of the phase transitions. Our EFM numerics show accurate agreement with previous predictions both on the location and on the nature of the transition thus unveiling the unifying non-local nature which characterizes all zero temperature phases with trivial and exotic \cite{hal} order.

\section{Model and phase diagram}
\begin{figure}[h!]
	\captionsetup[subfigure]{labelformat=empty}
	\centering
	\subcaptionbox{}
	{\includegraphics[scale=0.3]{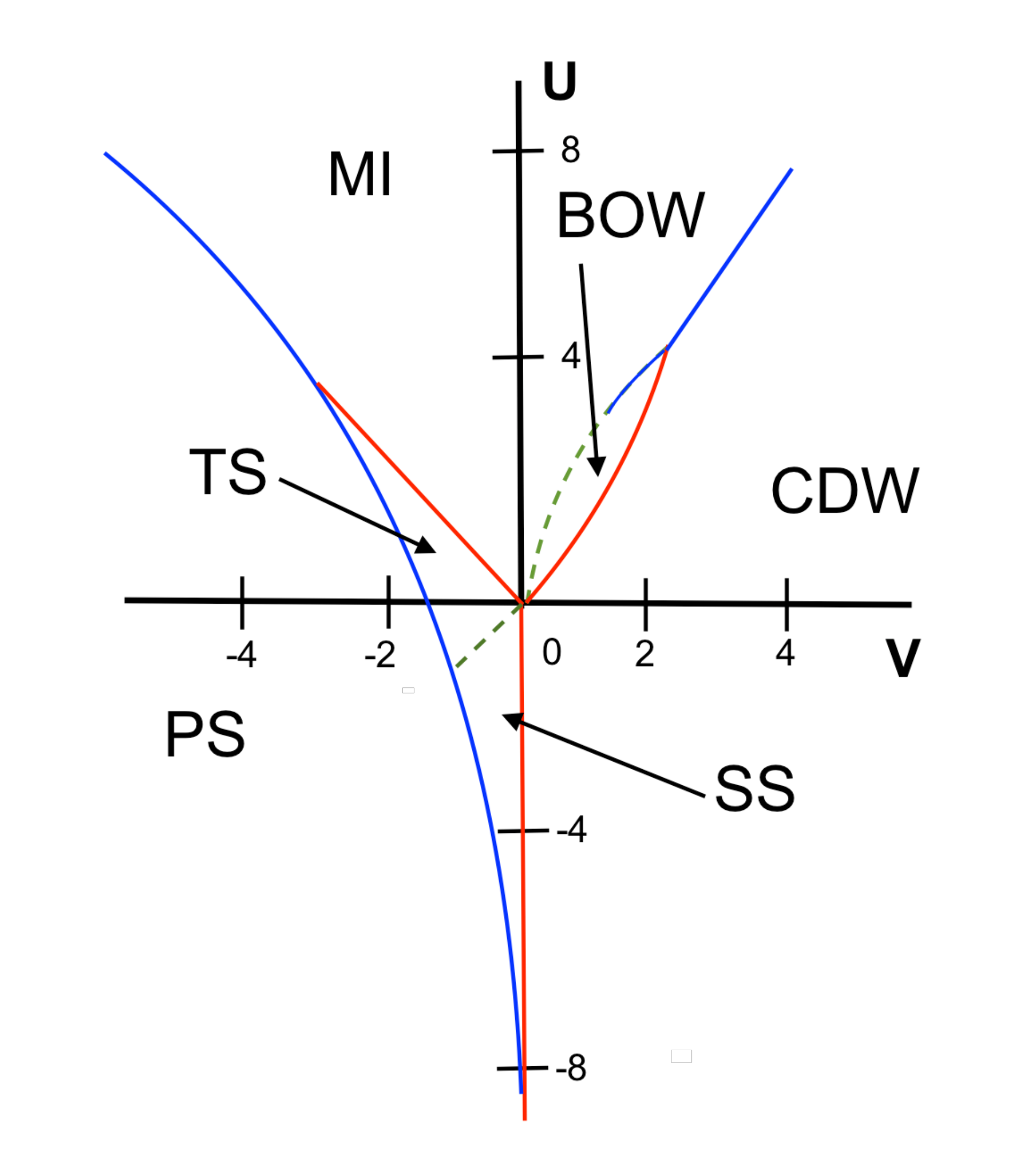}} 
	\hfill
	\subcaptionbox{}{\raisebox{2\height}{
			\begin{tabular}{llllll}
				\hline\noalign{\smallskip}
				&  $\tilde{\Phi_{c}}$ & $\tilde{\Phi}_{s}$ & $\Delta_{c}$ & $\Delta_{s}$ & LRO  \\
				\noalign{\smallskip}\hline\noalign{\smallskip}
				TS & $u$ & $u$ & 0 & 0 & none\\
				SS & $u$ & 0 & 0 & $\neq 0$ & $O_{P}^{s}$\\
				PS & - & - & 0 & $\neq 0$ & $O_{P}^{s}$\\
				MI & 0 & $u$ & $\neq 0$ & 0 & $O_{P}^{c}$\\ 
				BOW & 0 & 0 & $\neq 0$ & $\neq 0$ & $O_{P}^{c},\: O_{P}^{s}$\\ 
				CDW & $\frac{\pi}{2}$ & 0 & $\neq 0$ & $\neq 0 $& $O_{S}^{c},\, O_{P}^{s}$\\
				\noalign{\smallskip}\hline
			\end{tabular}
		}
	}
	\vspace{-2.3\baselineskip}
	\caption{{\it Left:} Cartoon of the phase diagram with $U$ and $V$ expressed in unit of $J$. Dashed green lines refer to KT transitions, blue lines to first order and red lines to continuous gaussian transitions. {\it Right:} Correspondence between ground state quantum phases and nonlocal operators that manifest long range order (LRO) \cite{BMR}. The letter \textit{u} indicates the unlocked fields and we have defined $\tilde{\Phi}_{\nu}=\sqrt{2\pi}\Phi_{\nu}$.}
\label{phasediag}	
\end{figure}

We consider a one dimensional unit density balanced two components Fermi mixture of $N$ fermions trapped in $L$ lattice sites, described by the following Hamiltonian
\begin{equation}
H=-J\sum_{j,\sigma}\left ( c_{j\sigma}^{\dagger}c_{j+1\sigma} + h.c.\right)
+U\sum_j n_{j\uparrow}n_{j\downarrow}+V\sum_{j}n_j n_{j+1} \quad ,
\end{equation}
where  $\sigma=\uparrow,\downarrow$ is the species index ($\bar{\sigma}$ denoting its opposite), $c_{j\sigma}^{\dagger}$ and $c_{j\sigma}$ are the fermionic creation and annihilation operators, respectively, $n_{j\sigma}= c_{j\sigma}^\dagger c_{j\sigma}$ counts the number of particles of species $\sigma$ and $n_j=\sum_{\sigma}n_{j\sigma}$. The coupling coefficients $J,U,V$ describe the tunneling processes, on-site and NN interaction respectively.  
The phase diagram, obtained in \cite{Hir,LiHi,CaFr,nakamura,lemini}, has a very rich structure, reported qualitatively in Fig.\ref{phasediag}. There spin-charge separation, characteristic of most low-dimensional fermionic systems, manifests in the independent opening of the charge and spin gaps $\Delta_\nu$ ($\nu=c,s$), defined respectively as $\Delta_c=(E(N+2,S_z=0)+E(N-2,S_z=0)-2E(N,S_z=0))/2$ and $\Delta_s=E(N,S_z=1)-E(N,S_z=0)$ being $E$ the ground state energy of a systems with $N$ fermions and total unbalance $S_z=N_{\uparrow}-N_{\downarrow}$ . Six different phases are present, identified by different dominant correlation functions. For strong $V>0$ usually a fully gapped charge density wave (CDW) phase characterized by alternating empty and doubly occupied sites takes place. On the other side, strong $U>0$ values support the presence of a charge gapped Mott insulator (MI) with uniform distribution of particles. Between the previous two phases for relatively weak repulsive interactions the competition between $U$ and $V$ generates a fully gapped bond ordered wave (BOW) as discussed above. Moreover, as it usually happens, for attractive strong on-site interaction fermionic pairing is achieved. In particular for $U<0$ and weak $V<0$ a spin gapped, conducting, singlet superconducting (SS) phase is present. Here the single particles are coupled in correlated pairs formed by up and down fermions in a background condensate of holons (empty sites) and doublons (doubly occupied sites). A strong attractive $V$ induces doublons to condense separately from holons, thus forming a spin gapped phase separated (PS) state. Finally, weak attractive $V$ and small $U$ support the formation of a fully gapless metallic regime with triplet superconducting (TS) order.
At low energies, the phase diagram can be also derived analytically by means of bosonization technique \cite{CaFr,Giam}, reproducing most of the above phases within the one-loop expansion. The Hamiltonian is mapped into two spin-charge decoupled sine-Gordon models, $H\rightarrow \sum_{\nu=c,s} H_\nu$, with
\begin{equation}\label{SG}
H_{\nu}=\frac{v_{\nu}}{2}\int dx \left[\left(\sqrt{K_{\nu}}\nabla\theta_{\nu}\right)^2+\left(\frac{\nabla\phi_{\nu}}{\sqrt{K_{\nu}}}\right)^2\right]+\frac{m_{\nu}v_{\nu}}{2\pi a^2}\int dx \cos{(\sqrt{8\pi}\phi_{\nu})}\; , \end{equation}
where the bosonic fields $\phi_\nu(x)$ are defined as customary, and the expression of the Luttinger parameters $K_\nu$, velocities $v_\nu$ and masses $m_\nu$ in terms of $J,U,V$ can be found in the literature \cite{Giam} (see also \cite{DoMo} for a review with the notation used here). In each channel, the competition between the kinetic and mass terms determines the regime(s) in which the latter becomes relevant, opening a gap. When this happens, depending on the sign of $m_\nu$, the field $\phi_\nu$ pins to $0$ or $\sqrt{\frac{\pi}{8}}$, ultimately producing a different physics of the gapped phase. In fact, for symmetry reasons, the spin field can only pin to the value $0$. The situation is resumed in the first four columns of the table in Fig. \ref{phasediag}. With the exception of the BOW \cite{note} and the PS phase, all other ordered phases which appear in Fig. \ref{phasediag} can be obtained in this way and are, for weak interactions, in agreement with the transition lines identified by numerical analysis.



\section{Non-Local Order Parameters}
Following \cite{dennijs,BDGA,MR,BMR,FMRB}, we introduce the parity (P) and string (S) non-local operators:
\begin{equation}
O_P^\nu(j)=\prod_{k=0}^{j-1}{\rm e}^{i\pi S_{z,k}^\nu} \quad ,\quad O_S^\nu(j)=\left (\prod_{k=0}^{j-1}{\rm e}^{i\pi S_{z,k}^\nu}\right )S_{z,j}^\nu \quad ,
\end{equation}
with $S_{z,j}^c=(n_j-1)$, $S_{z,j}^s=(n_{j\uparrow}-n_{j,\downarrow})$. In the asymptotic limit, their correlation functions $C_X^\nu=\lim_{r\rightarrow\infty}<[O_X^\nu(j)]^\dagger O_X^\nu(j+r)>$ ($X=P,S$) were claimed \cite{BMR} to remain finite each in correspondence of the opening of a specific gap in the $\nu$ channel. In particular within bosonization approximation $C_P^\nu\propto<(\cos\sqrt{2\pi} \phi_\nu)^2>$ remains finite when $\phi_\nu$ pins to the value $0$, whereas $C_S^\nu\propto<(\sin\sqrt{2\pi}\phi_\nu)^2>$ is finite when $\phi_\nu$ pins to the value $\sqrt{\frac{\pi}{8}}$. Thus the expectation value of $O_X^\nu$ configures as order parameter for the different gapped phases, as reported in Fig. \ref{phasediag}. Also, phases characterized by a non-vanishing $C_S^\nu$, see for instance \cite{FMRB}, are symmetry protected topological (SPT \cite{Wen}) phases \cite{MDIR}  with non trivial topological properties, like the presence of degenerate edge modes. Whereas phases with non-vanishing $C_P^\nu$ are trivial SPT phases. In addition to the topological properties, NLOPs are able to give crucial informations regarding the microscopic structure encoded in the ground state. Indeed, while a finite value of the parity related to the charge (spin) sector means that holon-doublon (up-down spin) virtual excitations are present, a non-zero charge (spin) string order signals the presence of hidden holon-doublon (up-down spin) antiferromagnetic order.
\section{Phase Transitions}
As shown in Fig. \ref{phasediag}, the phase digram of the EFM presents a large variety of phase transitions, namely KT, gaussian and first order. As shown for instance in \cite{FMRB}, this latter kind is usually not captured by standard bosonization approaches. For this reason we perform quasi-exact DMRG simulations which are efficiently able to detect any kind of phase transition. In the weak and intermediate $U,V>0$ regime it has been shown \cite{BMR} that NLOPs capture very accurately the phase diagram so we do not tackle this region. From the other side, it remains an open question whether NLOPs can give a proper description for different $U,V$ combinations. In our analysis we employ both periodic boundary conditions (PBC) and open boundary conditions (OBC). PBC are usually more suitable when NLOPs have to be evaluated in a computationally tough Hamiltonian, namely when the numerically accuracy can not be kept for large $L$. In particular, when PBC are used, boundary effects, which are quite relevant in the NLOPs extrapolation \cite{deb}, are removed and also relatively small $L$s allow to get a quite accurate extrapolation of the thermodynamic limit (TDL). From the other side, the PS regime makes the latter approach very challenging. This is due to the high energy degeneracy typical of this regime. For this reason, in all the phase transitions involving the PS regime, we used OBC, which allow to reduce the system degeneracy, while for all the other ones we employed PBC. More precisely, we extrapolated the TDL of the NLOPs from the finite size values $O_S^{\nu}(L/2)$ and $(O_P^{(\nu)}(L/2)+O_P^{(\nu)}(L/2+1))/2$. In the cases where PBC are employed, these were obtained for systems up to $L=32$ sites, keeping up to $1200$ DMRG states and performing 6 sweeps. From the other side, thanks to the higher DMRG efficiency, with OBC we studied system sizes with up to $L=56$ and a number of DMRG states ranging from $768$ to $1024$ with $5$ sweeps. Here, in order to minimize boundary effects, we cut out the first three sites in the evaluation of NLOPs.\\

\subsection{Attractive $U$ Regime}
\begin{figure}[h!]
	\centering
	\centering
	{\includegraphics[scale=0.4]{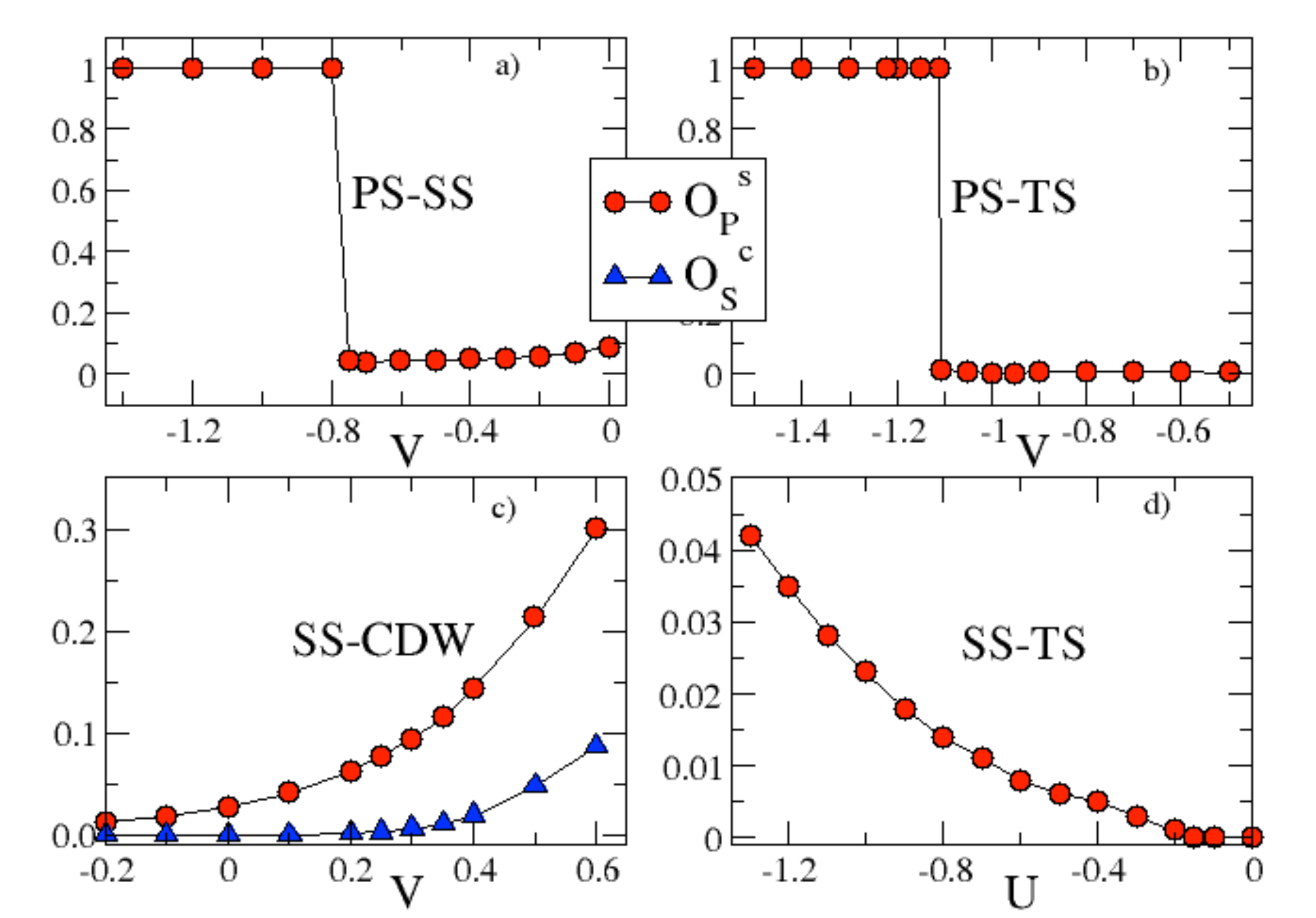}}
	\caption{TDL of the different NLOPs  with J=1 for: a) $U=-1.5$; b) $U=-0.1$; c) $U=-0.9$; d) $V=-0.1$.}
	\label{figneg}
\end{figure}  
As typical of fermionic lattice models, the $U<0$ regime is dominated by a trivial singlet-like spin gap associated with a finite value of $O_P^s$, as predicted by LL theory. In 1D, depending on the nature of particles involved, this kind of paring is usually associated to fermionic quasi-condensation (q-BEC) or superconducting (SC) order signaled by a power-law decay of the pair-pair correlation function. Here we show that $O_P^s$ acts as an order parameter also for the PS regime which, even if gapless in charge sector, has not any kind of q-BEC or SC behavior. As clearly visible in Fig. \ref{figneg} a) for the SS-PS transition, $O_P^s$ remains finite everywhere thus in agreement with the presence of a spin gap but its value is subject to a sudden jump when the transition is crossed. It signals a first order phase transition which is a crucial feature of transitions involving a PS regime. Indeed this is confirmed also for less attractive $U$. Here the competition between NN and on-site interaction makes possible a fully gapless TS regime while for more attractive $V$ a PS phase, with finite spin gap, is established. Here our results confirm again the presence of a first order transition associated to PS regime. Indeed Fig. \ref{figneg}b) shows again an abrupt jump of $O_P^s$ going from zero in the TS to almost one in the PS. From the other side, when a gapless-gapped transition not involving the PS phase takes place, as in the case of TS-SS in Fig. \ref{figneg}d), a slow continuos opening of  $O_P^s$ is observed. This is in agreement with the presence of an exponential opening of the gap typical of KT transitions as predicted in \cite{nakamura}. Finally, a gaussian continuos transition is observed in the charge sector once $V$ goes from attractive to repulsive values. Here, while the singlet-like pairing associated to $O_P^s$ remains finite, a charge gap associated to a CDW regime with charge antiferromagnetic order, thus with finite $O_S^c$, takes place, as shown in Fig. \ref{figneg} c).  

\subsection{Repulsive $U$ Regime}
\begin{figure}[h!]
	\centering
	\centering
	{\includegraphics[scale=0.4]{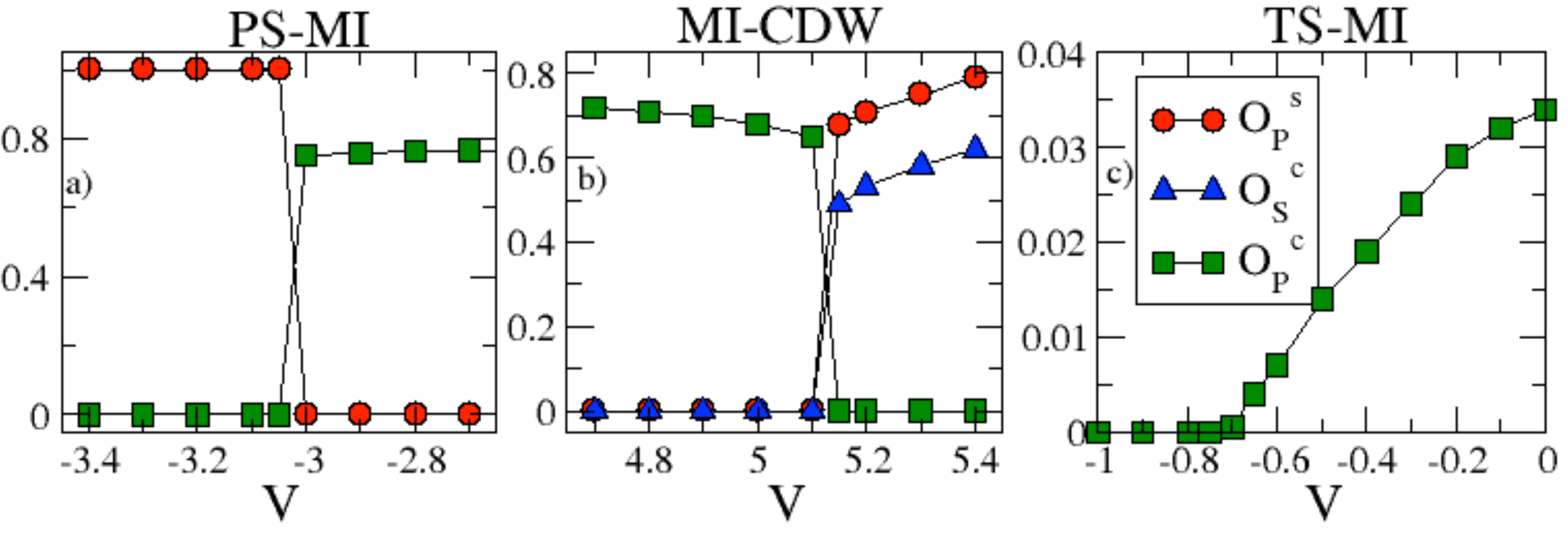}}
	\caption{TDL of the different NLOPs with J=1 for: a) $U=5.0$; b) $U=10$; c) $U=1$.}
	\label{figrep}
\end{figure}  
While the weak $U,V>0$ case has already been studied \cite{BMR}, the $U>0$, $V<0$ and the strong $U,V>0$ regimes have not been treated by means of NLOPs. For large $U$ and strongly attractive $V$ the systems is composed by two phases: the MI, with finite charge gap, and the PS, gapped in the spin sector. As already discussed in the previous section, usually it happens the phase transitions involving the PS regime are first order. This is again totally confirmed by our results in Fig. \ref{figrep} a) where clearly the two order parameters $O_P^c$ and $O_P^s$, for the MI and PS phase respectively, make an abrupt discontinuous jump from finite to zero values when the transition is crossed. Similar features can be observed in Fig. \ref{figrep} b). Here, even if the PS regime is not involved, the strong $V$ and $U$ considered make our system well inside the atomic limit which has strong analogies with a sort of classical regime. Indeed, similarly to the bosonic case, MI and CDW are separated by a first order transition. The latter manifests in a sudden opening of a spin gap signaled by $O_P^s$, combined to a change of the charge order once $V$ is increased. The situation is rather different when regimes with smaller coupling are studied. Indeed for weak $V<0$ a fully gapless TS regime appears, as previously shown also for attractive on-site interaction. Clearly, once U is increased, the strong quantum fluctuations are damped and an insulating regime, i. e. MI, is established. Here a gaussian transition occurs. It is signaled by the growing, from zero to finite values, of $O_P^c$, following the behavior shown in Fig. \ref{figrep} c).

\section{Conclusions}
We showed by numerical analysis how all phase transitions appearing in the extended Fermi-Hubbard model can be detected by means of non-local order parameters. In particular, while it was already seen that the latters work properly in transitions involving insulating states, our results show that also possibly conducting regimes with solely a spin gap are accurately captured. This provides a new perspective toward the detections of superconductivity or fermionic quasi-condensation. In addition our findings could have special impact on the experimental detection of phase transitions in one dimensional systems. Indeed, while most of the two-point correlation functions manifest phase transitions by a change in the decay law to zero, which is a very challenging experimental measure, our results show that non-local order parameters, being only zero or finite, might give a much more efficient description. Finally we stress that, since both our probes have already been measured \cite{endres} or are measurable \cite{parsons} and the extended Fermi-Hubbard Hamiltonian can be simulated within the ongoing experimental technologies involving magnetic atoms, our results could be tested in cold atomic experiments.


\begin{thebibliography}{}
	\bibitem{emery} V.J. Emery, S.A. Kivelson, and O. Zachar, Phys. Rev. B \textbf{56}, (1997) 6120.
	\bibitem{polimers} H. G. Keiss, \textit{Conjugated Conducting Polymers} (Springer-Verlag, Berlin, 1992)
	\bibitem{ishiguro} T. Ishiguro, K. Yamaji, \textit{Organic Superconductors} (Springer-Verlag, Berlin, 1990)
	\bibitem{Hir} J. E. Hirsch, Phys. Rev. Lett. {\bf 53}, (1984) 2327 
	\bibitem{LiHi} H. Q. Lin, J. E. Hirsch, Phys. Rev. B {\bf 33}, (1986) 8155 
	\bibitem{CaFr} J. W. Cannon, E. Fradkin, Phys. Rev. B {\bf 41}, (1990) 9435 
	\bibitem{PeMi} K. Penc, F. Mila, Phys. Rev. B {\bf 49}, (1994) 9670 
	\bibitem{nakamura} M. Nakamura, J. Phys. Soc. Jpn. {\bf 68}, (1999) 3123 ; M. Nakamura, Phys. Rev. B {\bf 61}, (2000) 16377
	\bibitem{lemini} F. Iemini, T. O. Maciel, R. O. Vianna Phys. Rev. B {\bf 92}, (2015) 075423 
	\bibitem{Tsuchiizu1} P. Sengupta, A. W. Sanders, D. C. Campbell, Phys. Rev. B {\bf 65}, (2002) 155113; M. Tsuchiizu, A. Furusaki, Phys. Rev. Lett. {\bf 88}, (2002) 056402 ; E. Jeckelmann, Phys. Rev. Lett. {\bf 89}, (2002) 236401 ; A. W. Sandvik, L. Balents, D. K. Campbell, Phys. Rev. Lett. {\bf 92}, (2004) 236401 ; M. Tsuchiizu, A. Furusaki, Phys. Rev. B {\bf 69}, (2004) 035103; S. Ejima, S. Nishimoto, Phys. Rev. Lett. {\bf 99}, (2007) 216403 ; M. Dalmonte, J. Carrasquilla, L. Taddia, E. Ercolessi, M. Rigol Phys. Rev. B {\bf 91}, (2015) 165136 
	\bibitem{BMR} L. Barbiero, A. Montorsi, M. Roncaglia, Phys. Rev. B {\bf 88}, (2013) 035109  
	\bibitem{dhar} A. Dhar, J. J. Kinnunen, P. T\"orm\"a, Phys. Rev. B {\bf 94}, (2016) 075116 
	\bibitem{Bloch2008} I. Bloch, J. Dalibard, W. Zwerger, Rev. Mod. Phys. {\bf 80}, (2008) 885 
	\bibitem{baier} S. Baier, M. J. Mark, D. Petter, K. Aikawa, L. Chomaz, Zi Cai, M. Baranov, P. Zoller, and F. Ferlaino,  Science {\bf 352}, (2016) 201-205 
	\bibitem{lu} M. Lu, N.Q. Burdick, S.H. Youn and B.L. Lev, Phys. Rev. Lett. {\bf 107}, (2011) 190401 
	\bibitem{aikawa} K. Aikawa,A. Frisch, M. Mark, S. Baier, R. Grimm, and F. Ferlaino Phys. Rev. Lett. {\bf 112}, (2014) 010404 
	\bibitem{park}J. W. Park, S. A. Will, M. W. Zwierlein Phys. Rev. Lett. {\bf 114}, (2015) 205302 
	\bibitem{DBRD} M.Di Dio, L. Barbiero, A. Recati, M. Dalmonte Phys. Rev. A {\bf 90}, (2014)  063608 
	\bibitem{FMRB} S. Fazzini, A. Montorsi, M. Roncaglia, L. Barbiero, arXiv:1607.05682
	\bibitem{endres} M. Endres et al., Science {\bf 334} , (2011) 200 
	\bibitem{parsons} M. F. Parsons, A. Mazurenko, C.S. Chiu, G. Ji, D. Greif, M. Greiner, arXiv:1605.02704; M. Boll, T. A. Hilker, G. Salomon, A. Omran, I. Bloch, C. Gross, arXiv:1605.05661
	\bibitem{mermin} N. Mermin, H. Wagner, Phys Rev. Lett. {\bf 17}, (1966) 1133 
	\bibitem{MDIR} A. Montorsi, F. Dolcini, R. Iotti, F. Rossi, arXiv:1610.0576v1
	\bibitem{dennijs} M. den Nijs and K. Rommelse, Phys. Rev. B {\bf 40}, (1989) 4709 
	\bibitem{white} S.R. White, Phys. Rev. Lett. {\bf 69}, (1992) 2863 
	\bibitem{hal}  F. D. M. Haldane, Phys. Rev. Lett.  {\bf 50}, (1983) 1153 
	\bibitem{Giam} T. Giamarchi, {\it Quantum Physics in One Dimension} (Oxford University Press, Oxford, 2003)
	\bibitem{DoMo} F. Dolcini, A. Montorsi, Phys. Rev. B {\bf 88}, (2013) 115115
	\bibitem{note} Notice that the BOW phase can be predicted by means of bosonization by considering also the second loop in the expansion, see \cite{nakamura}
	\bibitem{BDGA} E. Berg, E. G. Dalla Torre, T. Giamarchi, E. Altman, Phys. Rev. B {\bf 77}, (2008) 245119 
	\bibitem{MR} A. Montorsi, M. Roncaglia, Phys. Rev. Lett. {\bf 109}, (2012) 236404 
	\bibitem{Wen}  X. Chen, Z.-C. Gu, Z.-X. Liu, X.-G. Wen, Science {\bf 338}, (2012) 1604 ; Phys. Rev. B {\bf 87}, (2013) 155114 
	\bibitem{deb} C. Degli Esposti Boschi, A. Montorsi, and M. Roncaglia, Phys. Rev. B. 94, 085119 (2016).
	
\end{thebibliography}
\end{document}